\documentclass{emulateapj}                                                     

\usepackage{color}                

\shorttitle{Complete ionisation of the neutral gas in high redshift radio galaxies and quasars}
\shortauthors{Curran \& Whiting}

\def\ch2{$\chi^2$}

\def\Mo{M$_\odot$}

\def\ccm {$\hbox{{\rm cm}}^{-3}$}    

\def\MOLH {\hbox{${\rm H}_2$}}  

\def \AL {$\alpha $}     
\def \HI {H{\sc \,i}}

\def \WpHz {W Hz$^{-1}$}

\def\lapp{\ifmmode\stackrel{<}{_{\sim}}\else$\stackrel{<}{_{\sim}}$\fi}
\def\gapp{\ifmmode\stackrel{>}{_{\sim}}\else$\stackrel{>}{_{\sim}}$\fi}

\begin{document}


\title{Complete ionisation of the neutral gas: why there are so few detections of 21-cm hydrogen in high redshift
radio galaxies and quasars}


\author{S. J. Curran}
\affil{Sydney Institute for Astronomy, School of Physics, The University of Sydney, NSW 2006, Australia}
\affil{ARC Centre of Excellence for All-sky Astrophysics (CAASTRO)}

\and

\author{M. T. Whiting}
\affil{CSIRO Astronomy and Space Science, PO Box 76, Epping NSW 1710, Australia}
\email{sjc@physics.usyd.edu.au}

\begin{abstract}
  From the first published $z\gapp3$ survey of 21-cm absorption within the hosts of radio galaxies and quasars,
  \citet{cww+08} found an apparent dearth of cool neutral gas at high redshift.  From a detailed analysis of the
  photometry, each object is found to have a $\lambda = 1216$ \AA\ continuum luminosity in excess of $L_{1216}
  \sim10^{23}$ \WpHz, a critical value above which 21-cm has never been detected at any redshift.  At these wavelengths,
  and below, hydrogen is excited above the ground state so that it cannot absorb in 21-cm.  In order to apply the
  equation of photoionsation equilibrium, we demonstrate that this critical value also applies to the ionising ($\lambda
  \leq912$ \AA) radiation. We use this to show, for a variety of gas density distributions, that upon placing a quasar
  within a galaxy of gas there is {\em always} an ultra-violet luminosity above which all of the large-scale atomic gas
  is ionised.  While in this state the hydrogen cannot be detected nor engage in star formation.  Applying the
  mean ionising photon rate of all of the sources searched, we find, using canonical values for the gas density and
  recombination rate coefficient, that the observed critical luminosity gives a scale-length (3 kpc) similar that of the
  neutral hydrogen (\HI) in the Milky Way, a large spiral galaxy.  Thus, this simple, yet physically motivated, model
  can explain the critical luminosity ($L_{912} \sim L_{1216} \sim 10^{23}$ \WpHz), above which neutral gas is not
  detected.
This indicates that the non-detection of 21-cm absorption is not due to the sensitivity limits of current radio telescopes, but rather that
the lines-of-sight to the quasars, and probably the bulk of the host galaxies, are devoid of neutral gas.
\end{abstract}


\section{Introduction}
\label{intro}

Hydrogen gas accounts for 75\% of all the baryonic matter in the Universe, of which the cool component, the reservoir
for star formation, is traced by the radio-band 21-cm spin-flip transition. Due to the low probability of the transition,
compounded by the inverse square law, this is essentially undetectable at $z\gapp0.2$ (see \citealt{chg+08}), although in absorption
the line strength is dependent only upon the column density of the absorbing gas and the radio flux of the background source.

Hydrogen has been detected in the ultra-violet band Lyman-$\alpha$ transition, which traces all of the neutral gas, in
1500 high redshift galaxies {\em intervening} the sight-lines to more distant quasi-stellar objects (QSOs, see
\citealt{cwbc01,npls09}).  However, despite four decades of searches, knowledge of the cool component of this gas in the
distant ($z\gapp0.1$) Universe remains very scarce, with only 42 cases reported in these absorbers, intervening
radio-loud QSOs (quasars)\footnote{Compiled in \citet{cur09a}, with the addition of those recently reported by
  \citet{sgp+10,cwt+11}.}, in addition to 35 {\em associated} with the quasar host galaxy itself.\footnote{Compiled in
  \citet{cw10}, with the addition of three new associated absorbers, two reported in \citet{cwm+10} and one in
  \citet{cwwa11}. See also \citet{ace+12}.}

In both cases,
the majority of detections occur at redshifts of $z\lapp1$ (look-back times
$\leq7.7$ Gyr).\footnote{We employ a standard $\Lambda$ cosmology with $H_{0}=71$~km~s$^{-1}$~Mpc$^{-1}$, $\Omega_{\rm
    matter}=0.27$ and $\Omega_{\Lambda}=0.73$.}  
In the case of the intervening
absorbing galaxies, the apparent lack of cold gas at high redshift may be accounted for by geometry effects: In an expanding  Universe
 absorbers at redshifts of $z\gapp1$ are {\em always} disadvantaged, in comparison to the low redshift
($z\lapp1$) absorbing galaxies, in how effectively the absorber can cover the higher redshift background source \citep{cw06,cur12}.  

Since for
the associated systems the absorbing gas is located within the quasar host galaxy, such geometry effects cannot
account for the fact that the 21-cm detection rate at $z\lapp1$ is double that at $z\gapp1$. Furthermore, only one
associated 21-cm absorber has ever been found at $z>3$ \citep{ubc91}. This runs contrary to the expectation that at these
redshifts (look-back times $\geq11.5$ Gyr), much of the gas has yet to be consumed by star formation, meaning that we would
expect the abundance of hydrogen to be many times higher than in the present day Universe (e.g. \citealt{psm+01}).  

In addition to these covering factor effects,  for a given column density, the optical depth of the 21-cm absorption is dependent upon
the spin temperature of the gas \citep{wb75}. Since only atoms populating the lower hyperfine level can absorb in 21-cm,
the spin temperature may be elevated through:
\begin{enumerate}
  \item Excitation to the upper hyperfine level \citep{pf56}.
    \item Excitation above the ground state, particularly by  Lyman-\AL\ ($\lambda = 1216$ \AA) photons \citep{fie59}. 
      \item Ionisation.
\end{enumerate}
Although, with the data available, excitation to the upper hyperfine level through collisions (\citealt{pf56}, see also \citealt{dra11})
cannot be ruled out, \citet{cww+08} find no dependence of the 21-cm detection rate on the rest-frame 1420 MHz continuum
luminosity of the active galactic nucleus (AGN), thus at least ruling out excitation through this radiative process. 

\citet{cww+08} do, however, find a strong dependence on the rest-frame $\lambda = 1216$~\AA\ ($\nu = 2.47\times10^{15}$ Hz)
continuum luminosity. Specifically, that 21-cm absorption has {\em never} been detected above an apparent critical
luminosity of $L_{\rm 1216} \sim10^{23}$ \WpHz. For a $\approx50\%$ detection rate at $L_{\rm 1216}
\lapp10^{23}$ \WpHz\ \citep{cww+08}, the probability of 0 detections out of 19 searches occuring by chance is $1.9\times10^{-6}$
(significant at $4.76\sigma$ assuming Gaussian statistics, \citealt{cwm+10}). So although the gas may be excited through
other processes (collisions and the Cosmic Microwave Background), this correlation strongly suggests that excitation
above the ground state (and possible ionisation) by $\lambda \leq1216$ \AA\ photons is the dominant cause of the
non-detections.

Given that 17 of the 19 $L_{\rm 1216} \geq10^{23}$ \WpHz\ sources are type-1 AGN, 
it is also possible that the absorption, by cool gas in the circumnuclear obscuring torus invoked
by unified schemes, simply does not occur along our line-of-sight to the 
continuum source (e.g. \citealt{mot+01,pcv03,gss+06}). However, at $L_{\rm 1216} \leq10^{23}$ \WpHz, {\em both} type-1 and type-2
AGN exhibit a 50\% detection rate \citep{cww+08,cwm+10}, indicating the absorption must primarily arise
in the main galactic disk, which is randomly oriented with respect to the torus. Therefore, the bias towards type-1 objects at $L_{\rm 1216} \gapp10^{23}$ \WpHz\ is due to
these tending to arise in the more luminous quasars, as opposed to radio galaxies, which tend to be associated with type-2
objects and is therefore not an orientation effect \citep{cw10}.

Thus, \citet{cww+08} interpreted their exclusive non-detections at $z\gapp3$ to the traditional optical selection of targets,
in conjunction with the high redshifts, introducing a bias towards the sources more luminous in rest-frame ultra-violet.\footnote{Despite
shortlisting the faintest objects (with blue magnitudes of $B\gapp19$, see figure 5 of \citealt{cww09}).}
The same critical ultra-violet  continuum luminosity 
is also evident in the lower redshift surveys (see \citealt{ace+12}) and attributing
the lack of cold gas in the hosts of powerful AGN to the high ultra-violet luminosities
exciting the gas beyond detection (which we dub ``{\em the UV interpretation}''), can explain why this effect is seen at all
redshifts. The UV interpretation may also account for several other issues in extragalactic radio astronomy, such as the
elevated detection rate in compact objects and the
preference for 21-cm detection in radio galaxies over quasars \citep{cw10}.

Given the low probability of zero detections occuring by chance above a given $1216$ \AA\  continuum luminosity,
in conjunction the fact that $\lambda < 1216$ \AA\ photons
excite (and possibly ionise) the gas so that it cannot absorb in 21-cm, there is little doubt that the UV
interpretation is the correct physical description. This has been confirmed by an independent survey
for 21-cm in 143 radio sources at redshifts $0.02 < z < 3.8$, where the lack of detections is correlated with the UV luminosity \citep{gd11},
as well as by \citet{psv+12}, who find a critical X-ray luminosity, above which sources are not detected in 250 $\mu$m continuum emission,
a tracer of star formation.

However, one
question remains unanswered: Why is there a hard limit to the UV luminosity, above which the gas is excited beyond
detection by the most sensitive radio telescopes, rather than a continuum where the detections
gradually become fewer and fewer as the ultra-violet luminosity increases? We address this issue here.

\section{Photoionsation Equilibrium}

For a cloud of hydrogen containing an ionising source, the equilibrium between photoionsation and recombination of
protons and electrons in a nebula can be written as \citep{ost89},
\begin{eqnarray}
  \int^{\infty}_{\nu_{_{\rm ion}}}\frac{L_{\nu}}{h\nu}\,d{\nu}= 4\pi\int^{r_{\rm ion}}_{0}\,n_{\rm p}\,n_{\rm e}\,\alpha_{A,B}\,r^2\, dr ,
\label{eq1}
\end{eqnarray}
where $L_{\nu}$ is the specific luminosity at frequency $\nu$ and $h$ is the Planck constant, giving the number of
ionising photons per second. On the right hand side, 
$r_{\rm ion}$ is the extent of the ionisation,
$n_{\rm p}$ and  $n_{\rm e}$  are the proton and electron densities, respectively,  and $\alpha_{A,B}$
 the radiative recombination rate coefficient of hydrogen (see Sect.~\ref{sect:models}). 

\begin{figure*}[h]
\centering
\includegraphics[angle=-90,scale=0.73]{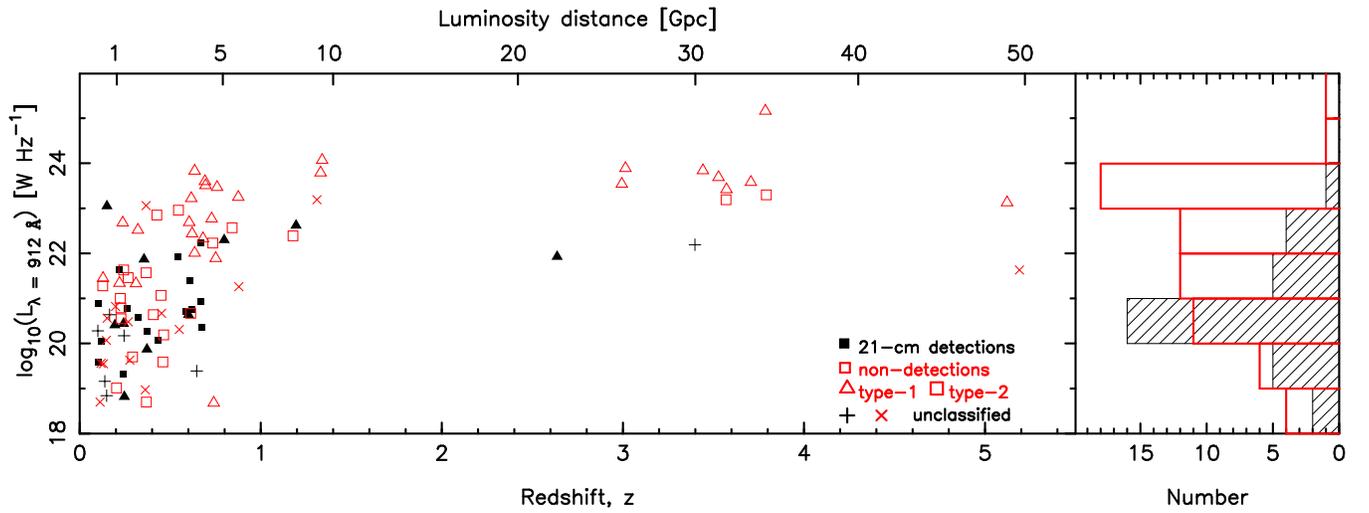}
\caption{The $\lambda = 912$ \AA\ continuum luminosity (where available, see \citealt{cwsb12})
versus redshift for the sources searched in 21-cm absorption. The filled symbols/hatched histogram represent the 21-cm
detections and the unfilled symbols/unfilled histogram the
non-detections. The shapes represent the AGN classifications, with
triangles representing type-1 objects and squares type-2s ({\bf +} and
{\sf x} designate an undetermined AGN type for a detection and
non-detection, respectively). 
}
\label{912-z}
\end{figure*}

Since, after excitation to the upper hyperfine level, the next excitation is to ${\sf n} = 2$
 by Lyman-\AL\ photons, our proxy has been the $\lambda = 1216$~\AA\ continuum luminosity.
However, since excitation to the ${\sf n}=2$ level and ionisation of the hydrogen atom are so close in energy (both events being
  $\approx2\times10^6$ times as energetic as the spin-flip transition), this critical luminosity
should also apply in the case of ionisation. In order to verify this, in Fig.~\ref{912-z} we show the $\lambda = 912$~\AA\ continuum luminosity
distribution.
The luminosities have been derived from the photometries as described in \citet{cww+08}, but with the 
inclusion of data from the Galaxy Evolution Explorer (GALEX, \citealt{mbb+03}). These, in conjunction with the 
$BVRK$ magnitudes from the literature, allow reliable power-law fits to the rest-frame UV data 
(corrected for Galactic extinction using the maps of
\citealt{sfd98}) over a range of
redshifts, from which the $\lambda = 912$ \AA\ continuum luminosities were derived (see \citealt{cwsb12}).

  
From Fig. \ref{912-z}, we see that the same approximate critical value applies in the case of ionising
photons. That is, 21-cm absorption has never been detected above a luminosity close to $L_{912} \sim10^{23}$ \WpHz. 
The largest measured 912 \AA\ luminosity for which there is a detection is $L_{912} = 1.1\times10^{23}$ \WpHz,
above which there are 20 non-detections. 
Of the sources for which we could reliably determine $L_{912}$, there are 38 detections
and 60 non-detections (i.e. a 39\% detection rate) below 
this luminosity. Applying this probability of $p=0.61$ for a non-detection to the $L_{912} > 1.1\times10^{23}$ \WpHz\ sources, there is
a  binomial probability of $5.09\times10^{-5}$ of the 20 non-detections occuring by chance, a $4.05\sigma$ significance.

Thus, although only excitation above the ground state is required to explain the dearth of 21-cm
absorption in UV luminous sources, it is possible that ionisation of the gas is the primary
cause of the non-detections. Given that the lifetime in the ${\sf n}=2$  state is
only $\sim10^{-8}$~sec, this is the more likely situation and so we are justified in applying
Eq. \ref{eq1} to this problem.
Thus, in Sect. \ref{sect:seds} we derive the value of the left hand side of the equation, for the sources
searched in 21-cm, and in Sect. \ref{sect:models},
we apply various 
recombination models to the right hand side of the equation.

\subsection{Photoionsation rates}
\label{sect:seds}

In addition to determining the UV fluxes from the GALEX photometries and the $BVRK$ magnitudes, in order to investigate
differences between the 21-cm detected and UV luminous non-detected samples, we obtained multi-wavelength data from all
of the relevant photometries given by the NASA/IPAC Extragalactic Database (NED), again correcting for Galactic
extinction using the maps of \citet{sfd98}.  As per the $\lambda = 912$~\AA\ continuum luminosities, above, we used the
redshifts of the targets to blue-shift the observed frequencies back to the source rest-frame values and converted the
observed fluxes to luminosities. We then averaged all of the luminosities within a specified frequency range to obtain a
composite SED for all of redshifted the sources searched in 21-cm absorption.

\begin{figure*}[h]
\centering
\includegraphics[angle=0,scale=0.9]{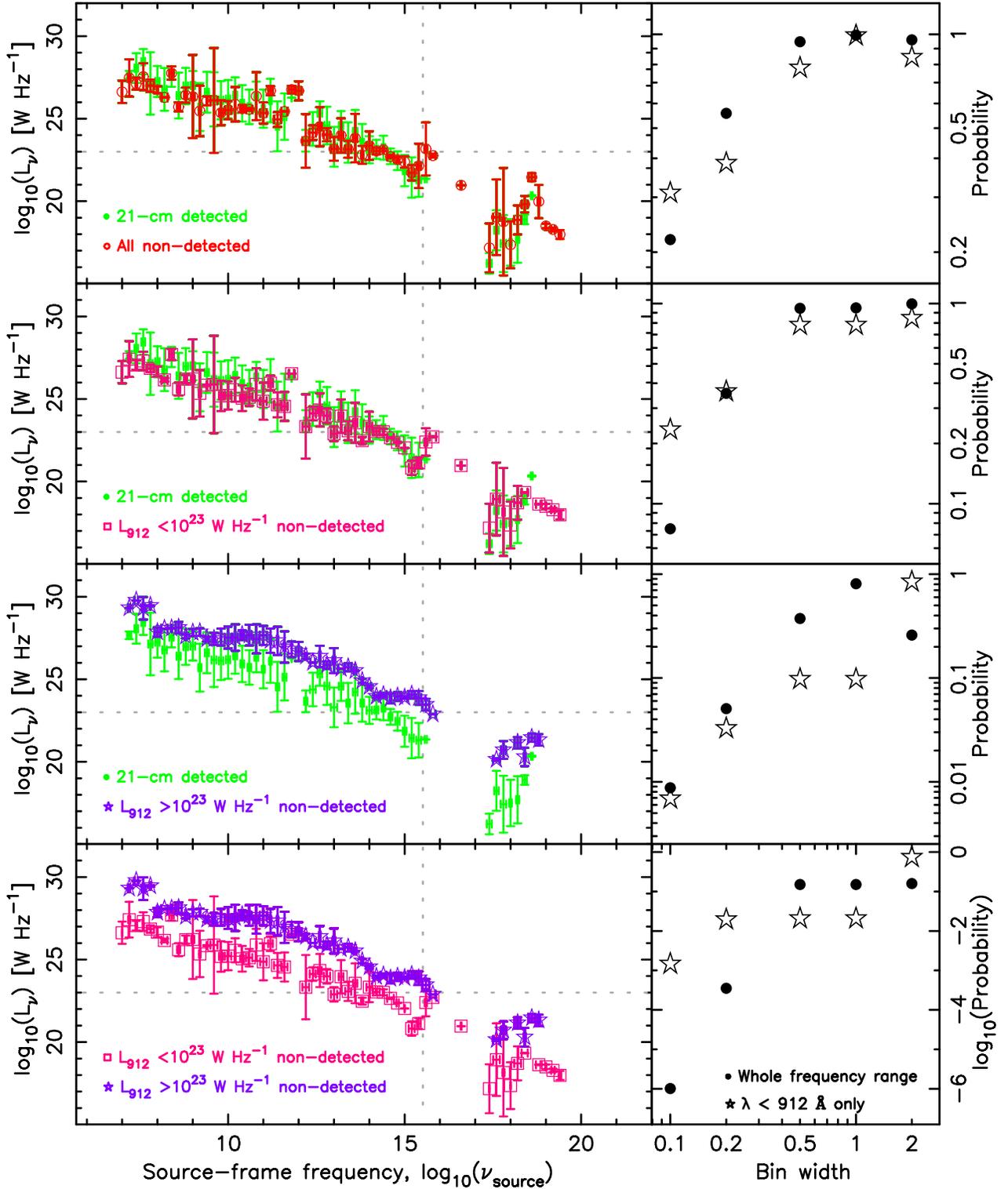}
\caption{The composite SEDs for each of the various sub-samples for a bin width of $\pm0.1$ (i.e.
each bin is $10^{2\times0.1} = 1.6$ times the frequency of the previous). In the left panel 
the vertical dotted line shows the source-frame frequency of $\nu = 3.29\times10^{15}$ Hz  ($\lambda=912$ \AA),
above which the photons ionise the atom. The right panel shows the probability, from
a Kolmogorov-Smirnov test, that the two samples tested are from the same population for various bin widths
for both the whole frequency range (filled circles) and  $\lambda\leq 912$ \AA\ (hollow stars).}
\label{SED_0.1}
\end{figure*}
Performing a Kolmogorov-Smirnov test between the binned luminosities of the various sub-samples (shown in each panel of
Fig. \ref{SED_0.1}), we find no evidence that the 21-cm detected and $L_{\rm UV}\lapp10^{23}$ \WpHz\ non-detected samples
are drawn from different populations, with a probability of $\gapp0.2$ (for all bin widths) that they are drawn from
the same population. However, between either the
21-cm detected/$L_{\rm UV}\lapp10^{23}$ \WpHz\ non-detected samples and the $L_{\rm UV}\gapp10^{23}$ \WpHz\ sample, it
is seen that, for sufficiently high resolution bins, the probability can get as low as $1\times10^{-6}$, although this
is due to the extra high frequency points in the $L_{\rm UV}\lapp10^{23}$ \WpHz\ non-detected sample, with a probability
of $\approx0.004$ being more likely. This still suggests, however, that the UV luminous sources are drawn from a
different sample than those with $L_{\rm UV}\lapp10^{23}$ \WpHz.

\begin{figure*}[h]
\centering
\includegraphics[angle=0,scale=0.9]{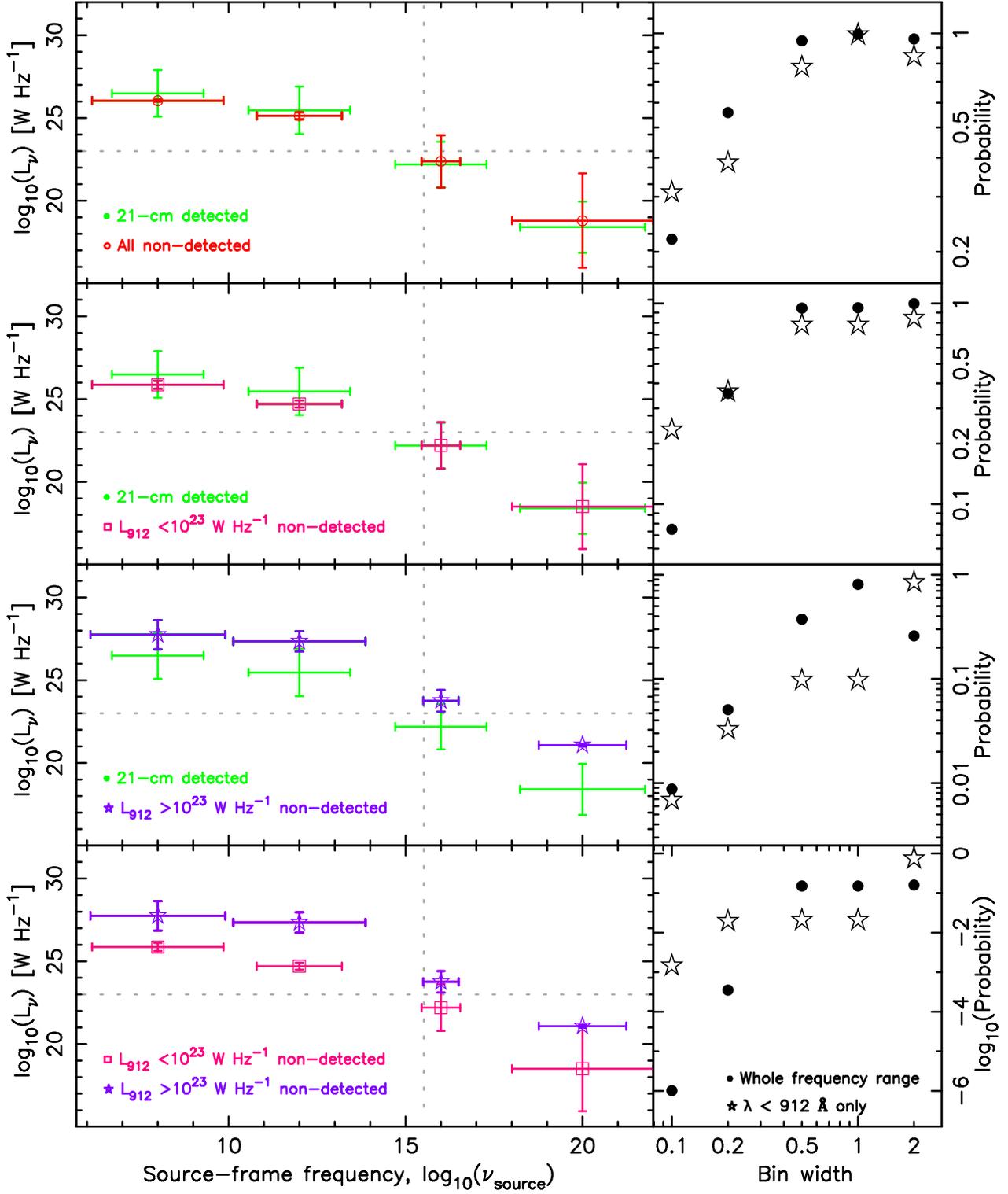}
\caption{As Fig. \ref{SED_0.1} but for a bin width $\pm2$  (i.e.
each bin is $10^{2\times2} = 10\,000$ times the frequency of the previous). In this and Fig. \ref{SED_0.1}, the symbol
shows the mean luminosity at the centre of the frequency bin (e.g. at $10^8,\,10^{10},...\,10^{20}$ Hz here)
with the errors bars showing the standard deviation in both luminosity and frequency.}
\label{SED_2.0}
\end{figure*}

In order to obtain the rate of ionising photons, we are interested in frequencies above $\nu = 3.29\times10^{15}$ Hz. However,
as seen from Fig. \ref{SED_0.1}, there is a large gap in the SEDs between $\sim10^{16}$ and $\sim10^{17}$ Hz, the range
of spaced-based ultra-violet observations between the optical and X-ray bands. Although the X-ray
observations are also space-based, these generally have more sky coverage than the ultra-violet observations and are thus more likely to have
observed one of our sources. So, in order to obtain an estimate of $\int^{\infty}_{\nu_{\rm ion}}
(L_{\nu}/\nu)\,d\nu$, where $\nu_{\rm ion}= 3.29\times10^{15}$ Hz, we smooth the SEDs (Fig. \ref{SED_2.0}) and
interpolate a power-law fit between $\sim10^{16}$ and $\sim10^{20}$ Hz
to obtain the mean dependence of $L_{\nu}$ on $\nu$.
The photon rate is given by
\[
\int^{\infty}_{\nu_{_{\rm ion}}}\frac{L_{\nu}}{h\nu}\,d{\nu},~{\rm where}~\log_{10}L_{\nu} = \alpha\log_{10}\nu+ {\cal C} \Rightarrow  L_{\nu} = 10^{\cal C}\nu^{\alpha}
\]
for a power-law, where \AL\ is the spectral index and ${\cal C}$ the intercept. Solving this,
\[
\frac{10^{\cal C}}{h}\int^{\infty}_{\nu_{_{\rm ion}}}\nu^{\alpha-1}\,d{\nu} = \frac{10^{\cal C}}{\alpha h}\left[\nu^{\alpha}\right]^{\infty}_{\nu_{_{\rm ion}}} = \frac{-10^{\cal C}}{\alpha h}\nu_{_{\rm ion}}^{\alpha}~{\rm where}~\alpha < 0.
\]
From the composite SEDs, for the non-UV luminous sample we find $L_{\nu}\approx10^{37.3}\,\nu^{-0.95}$ \WpHz, giving
$5.5\times10^{55}$ ionising photons~sec$^{-1}$ and for the UV luminous sample, $L_{\nu}\approx10^{34.6}\,\nu^{-0.68}$
\WpHz, giving $2.9\times10^{57}$ ionising photons~sec$^{-1}$. This is $\approx50$ times the luminosity of the 21-cm
detected sample, which is consistent with the factor of $\approx7$ in the luminosity distances between the $z\gapp3$
sample and the cluster of 21-cm detections at $z\lapp0.9$ (Fig. \ref{912-z}).

\subsubsection{The critical photoionsation rate}

Since we are interested in the ionising photon rate resulting from a critical luminosity of $L_{\rm UV}\sim10^{23}$ \WpHz\
(Fig. \ref{912-z}), we use the highest 21-cm detected luminosity of $L_{\rm 912} = 1.1\times10^{23}$ \WpHz\ at
$3.29\times10^{15}$ Hz, together with the above spectral index of $\alpha=-0.68$, to obtain
$L_{\nu}\approx10^{33.6}\,\nu^{-0.68}$ \WpHz, which gives $2.9\times10^{56}$ sec$^{-1}$ for the critical ionising photon
rate.  Referring to the literature, from the spectra of several hundred QSOs, \citet{tzkd02} find a mean optical--X-ray
slope of $\alpha=-1.5$. This is significantly steeper than the mean spectral index derived for our sample, which
consists exclusively of powerful radio sources, although applying a critical luminosity of $L_{\rm 912} = 1.1\times10^{23}$ \WpHz\
gives  $L_{\nu}\approx10^{46.3}\,\nu^{-1.5}$ \WpHz\ $\Rightarrow 1.1\times10^{56}$ sec$^{-1}$, which is in the ballpark of the
value derived for our sample. This is a consequence of the steeper spectral index being compensated by a larger constant
(intercept) and the fact that the lower frequency end of the UV SED ($\nu\sim3.3\times10^{15}$ Hz) contains most of the
energy. We are therefore confident in applying $\int^{\infty}_{\nu_{\rm ion}}\ (L_{\nu}/h\nu)\,d\nu = 2.9\times10^{56}$
photons sec$^{-1}$ to the left hand side of Eq. \ref{eq1}.\footnote{Other studies of the UV continuum slope in 
quasars and AGN indicate their spectral indices to be in the range $\alpha=-1.4$ to $-0.56$, although with
considerable scatter. These give 
critical photon rates of  $\sim5\times10^{55}$ and  $\sim2\times10^{54} -  2\times10^{57}$ sec$^{-1}$
(\citealt{skb+04,ssd12}, respectively).}

\begin{figure*}[h]
\centering
\includegraphics[angle=-90,scale=0.73]{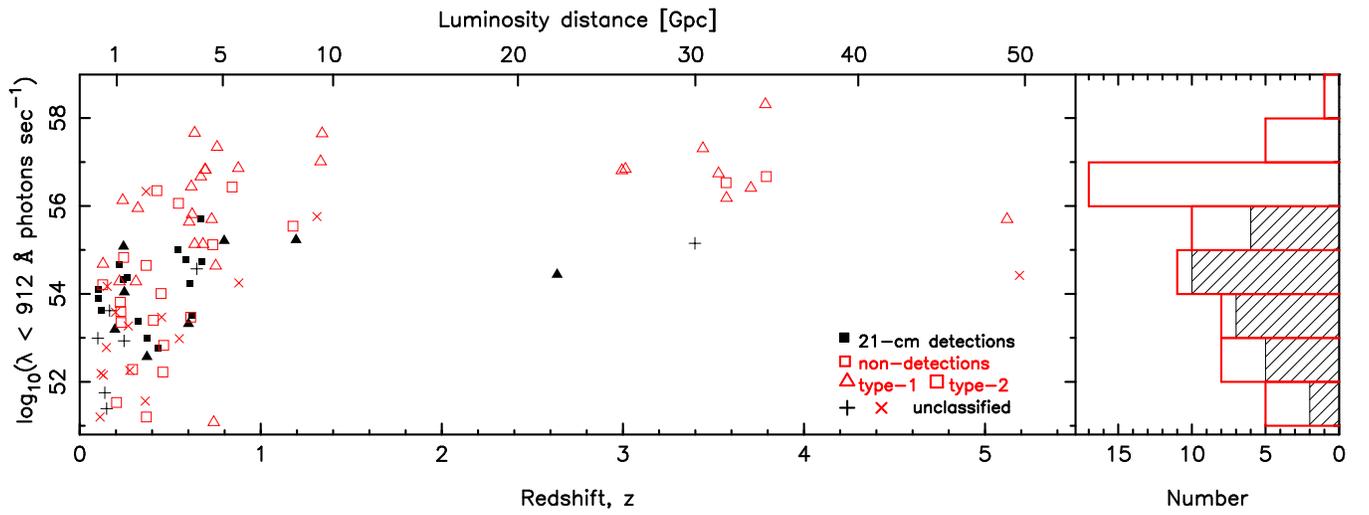}
\caption{The rate of ionising ($\lambda < 912$ \AA, bottom) photons
versus redshift for the sources searched in 21-cm absorption, obtained from the polynomial fits to the SEDs  \citep{cwsb12}. 
The symbols and histograms are as per Fig. \ref{912-z}.
}
\label{ion-z}
\end{figure*}

Lastly, it is clear that, while the specific continuum luminosity ($L_{1216}$ or $L_{912}$) may provide an indicator
of the amount ionising radiation from the AGN, the integrated ionising luminosity (i.e. the ionising photon rate)
is the correct measure. By fitting polynomials to the photometry of the individual sources \citep{cwsb12}, 
in Fig.~\ref{ion-z} we
re-plot Fig.~\ref{912-z} in terms of  $\int^{\infty}_{\nu_{\rm ion}} (L_{\nu}/h\nu)\,d\nu$.
For those for which accurate polynomial fits could be obtained,  we see that above $\sim10^{56}$ ionising photons~sec$^{-1}$, 21-cm searches have
resulted in exclusive non-detections. The highest photon rate for a detection, which could be reliably determined,  
is $1.7\times10^{55}$~sec$^{-1}$, a value above which there are 29 non-detections.  
Below this rate there are 38 detections and 51 non-detections,
giving a $p=0.57$ probability of a non-detection. Using this proxy, the 
binomial probability of 29 out of 29 non-detections occuring by chance is just $8.32\times10^{-8}$. This $5.36\sigma$ 
result therefore strongly
suggests that ionisation of the gas by $\lambda\leq 912$~\AA\ photons from the AGN is responsible for the non-detection
of 21-cm absorption in high redshift sources.

\subsection{Recombination models}
\label{sect:models}

We now parametrise the right hand side of the photoionsation equilibrium expression (Eq. \ref{eq1}).  Since we are
concerned with the ionisation of neutral gas and its subsequent recombination, $n_{\rm p} = n_{\rm e} = n$.
Also, in optical band observations of an optically thick plasma, where
direct capture onto the ground state is excluded, $\alpha_{\rm B}$ is used. However, since we are concerned with the
ground state, $\alpha_{\rm A}$ is the relevant total recombination rate coefficient \citep{of06}.  We choose this value
at $2000$~K, $\alpha_{\rm A} = 1.27\times10^{-12}$ cm$^3$ sec$^{-1}$,\footnote{Compared with $\alpha_{\rm B} =
  0.90\times10^{-12}$ cm$^3$ sec$^{-1}$, hence the choice of $\alpha_{\rm A}$ or $\alpha_{\rm B}$ making little difference.} the typical upper limit to the spin
temperature found in intervening absorbers (when the Lyman-\AL\ line is also detected and an upper limit to the spin
temperature can be determined, \citealt{ctd+09}).

Na\"{i}vely assuming a constant particle density of $n = 10$~\ccm\ (typical of the cool neutral 21-cm absorbing
interstellar medium) throughout the nebula, we find $r_{\rm ion} = 3$ and $13$~kpc for the mean radii of the ``Str\"{o}mgren
spheres'' of the UV non-luminous and UV luminous samples, respectively.  Although the latter value is of the same order
of magnitude as the extent of neutral gas in a large galaxy, this model represents a gradual increase in ionised radius
with luminosity, with no critical value.

\subsubsection{Exponential gas density distribution}

\begin{figure}[h]
\centering \includegraphics[angle=270,scale=0.73]{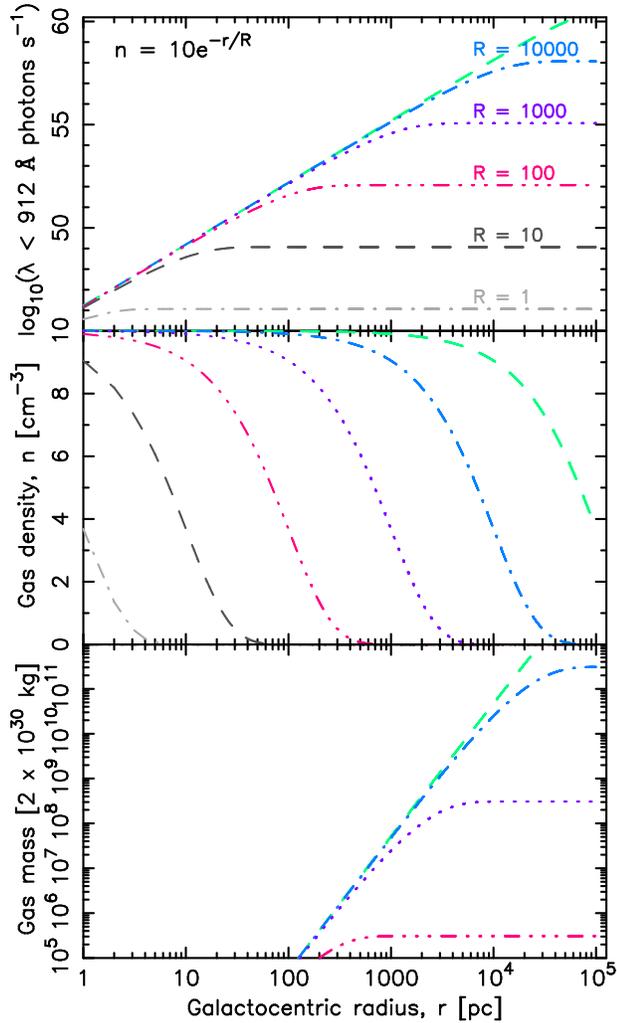}
\caption{The number of ionising  ($\lambda < 912$ \AA) photons per second (top), particle density (middle) and gas
  mass in solar masses (bottom) versus the galactocentric radius for an exponential gas distribution at a temperature of
  $2000$ K ($\alpha_{\rm A} = 1.27\times10^{-12}$ cm$^3$ sec$^{-1}$).  The different line styles represent
  the various scale-lengths, $R$, in parsecs, applied to the gas density distribution, $n = n_0\, e^{-r/R}$, with the
  top panel showing the radius of ionised gas for each value of $R$.}
\label{X-n}
\end{figure}

Unlike the idealised ionised region around a star, we do not expect the gas density to remain constant on galactic
scales.  A more realistic model of the density of the cold neutral medium (CNM) within a galaxy is that of an
exponential decrease in the gas density with distance from the nucleus \citep{bbs91,kdkh07}.  Thus, for $n = n_0\,
e^{-r/R}$, where $n_0$ is the gas density at $r = 0$ and $R$ is a scale-length describing the rate of decay of this with
radius, 
Eq. \ref{eq1} becomes
\begin{eqnarray}
\int^{\infty}_{\nu_{_{\rm ion}}}\frac{L_{\nu}}{h\nu}\,d{\nu} = 4\pi\,\alpha_{\rm A}\,n_0^2\int^{r_{\rm ion}}_{0}\,e^{-2r/R}\,r^2\, dr \nonumber \\ = \pi\,\alpha_{\rm A}\,n_0^2\left[R^3- R\,e^{-2r_{\rm ion}/R}\left(2\,r_{\rm ion}^2 + 2\,r_{\rm ion}R + R^2\right)\right].
\label{eq2}
\end{eqnarray}
Unlike the constant density distribution, this
becomes independent of $r$ at sufficiently large radii, i.e. $\int^{\infty}_{\nu}\
(L_{\nu}/h\nu)\,d\nu \rightarrow  \pi\,\alpha_{\rm A}\,n_0^2\,R^3 $. Conversely, for a given scale-length,
$R$, there {\em always} exists a ``ceiling luminosity''  (the number of ionising photons $\times \,h$) 
for which {\em all} of gas is ionised (Fig.~\ref{X-n}).

Using the above values of $\alpha_{\rm A}=1.27\times10^{-12}$ cm$^3$ sec$^{-1}$ and $n_0 = 10$ cm$^{-3}$, the critical ionising
photon rate of $2.9\times10^{56}$ sec$^{-1}$ gives a scale-length of $R=2.9$ kpc.
We can compare this to the \HI\ in the Milky
Way, where \citet{kk09}  fit an exponential profile to the mid-plane
volume density distribution to find 
$R=3.15$~kpc and $n_0 = 0.9\,e^{R_\odot/R} = 13.4$~cm$^{-3}$. This is in close agreement with our values,
demonstrating that the mean SED normalised by a $\lambda = 912$~\AA\ continuum luminosity of $10^{23}$ \WpHz\ is sufficient to 
ionise all of atomic gas in a large spiral galaxy, rendering it  undetectable in 21-cm.

It is therefore clear that an exponential decrease in gas density with distance from the nucleus can naturally yield a
critical value in the UV luminosity which is close to that found observationally. This does however, rely on a simple
model of the CNM, within which various structures and phases will be embedded, such as the warm
neutral medium, as well as localised regions of ionised gas and dense molecular clouds. However, here we are modelling the
large-scale CNM, for which an exponential density distribution is a realistic model \citep{bbs91,kdkh07}. 

Although using the canonical values for $\alpha$ and $n_0$ gives the correct
scale-length for the observed photon rate, a further physical (sanity)
check can be obtained by deriving the total gas mass from the gas density and volume via $M_{\rm gas} =
\int_0^{r}\,\rho\, dV$. In this case, where the particle density of $n$ protons cm$^{-3}$ corresponds to $\rho =
1.67\times10^{-21}\times n$ kg m$^{-3}$, exponentially decaying with $r$ across a disk of thickness, $t\,(r) = r/f_{\rm
  FL}$, we have
\begin{eqnarray}
M_{\rm gas} = 2\pi\,n_0\,\int_0^{r}\,e^{-r/R}\,r\, t \,dr = \frac{2\pi\,n_0}{f_{\rm FL}}\int_0^{r}\,e^{-r/R}\,r^2 \nonumber \\ =\frac{2\pi\,n_0\,R^3 }{f_{\rm FL}}\left[2 - e^{-r/R}\left(\frac{r^2}{R^2} + 2\frac{r}{R} +2\right)\right],
 \label{eq3}
\end{eqnarray}
where the flare factor, $f_{\rm FL}$, describes the flaring of the \HI\ gas scale-height with galactocentric radius.
Applying the mean Milky Way value of $f_{\rm FL}\approx20$ \citep{kdkh07}, a scale-length of $R=2.9$ kpc gives a total gas mass of
$M_{\rm gas} =7.5\times10^{9}$ \Mo\ (Fig.~\ref{X-n}, bottom panel).
This is close to the mean value found from a low redshift survey of 21-cm emission from the 1000 \HI\ brightest
galaxies in the southern sky \citep{ksk+04}, giving us further confidence in the exponential decay model and choice of gas density.

\subsubsection{Alternative temperatures and disk profiles}

\begin{figure}[h]
 \centering 
 \includegraphics[angle=270,scale=0.73]{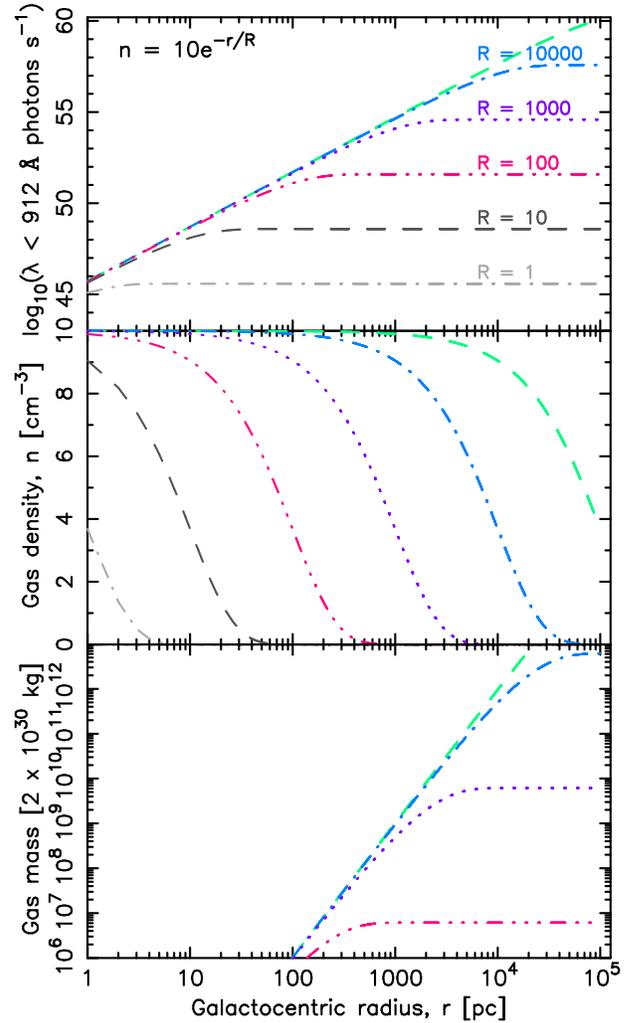}
 \caption{As Fig. \ref{X-n} but for a gas temperature of $10\,000$ K ($\alpha_{\rm A} =4.19\times10^{-13}$ cm$^3$ sec$^{-1}$) and a spherical gas distribution.}
 \label{X-n-4}
\end{figure}

Although spin temperatures in intervening 21-cm absorbers may be, on average, $\lapp2000$~K, without the total neutral
hydrogen column density from the Lyman-\AL\ transition, as is the case for the associated absorbers, an upper limit to
the spin temperature cannot be computed. Traditionally in the discussion of gas ionisation, temperatures of $\sim10^4$ K
are brandished (e.g. \citealt{ost89,hr01}), although these are in the case of Lyman-\AL\ emission, rather than the much
less energetic 21-cm absorption. Given that the gas is most likely ionised however, in Fig.~\ref{X-n-4} (top panel) we show
the ionising luminosity versus the extent of the ionised gas for $\alpha_{\rm A} = 4.19\times10^{-13}$ cm$^3$ sec$^{-1}$
(i.e. at $10^4$ K ). From the observed critical rate of $2.9\times10^{56}$ photons sec$^{-1}$, we see that the
scale-length increases to $R=4.2$ kpc, which is larger than that of the Milky Way.
Thus, for this temperature the photon rate is more than that required to ionise
all of the neutral gas, while demonstrating that our
main result is not overly sensitive to the choice of temperature.\footnote{Since $\alpha\propto\sqrt{T}$ (http://amdpp.phys.strath.ac.uk/tamoc/DATA/RR/)} 

Although having no effect on the extent of the ionisation, 
in order to investigate the effect of a different disk profile on the mass, in Fig.~\ref{X-n-4} we show
the distribution of gas mass for a sphere, rather than a disk (i.e. $dV = 4\,\pi\,r^2\,dr$).  For
$R=4.2$ kpc, this gives a total gas mass of $M_{\rm gas} =4.6\times10^{11}$ \Mo, which is close to the total (dynamical)
mass expected in a galaxy and is thus too high\footnote{See \cite{ckb08} for an inventory of the various masses in a
  near-by active galaxy.} and, applying a temperature of $2000$~K (i.e. $R = 2.9$ kpc), lowers this only slightly to $M_{\rm gas}
= 1.5\times10^{11}$ \Mo. This confirms that the disk model, which reproduces a gas mass close to the typically observed
value, is the more physically accurate distribution. Furthermore, although the most luminous quasars may reside in
elliptical galaxies \citep{tdhr96}, this shape traces the stellar distribution and not necessarily that of the neutral
gas. For instance, ``superdisks'' of gas and dust in the elliptical hosts of powerful radio galaxies have been proposed
\citep{akmv98,gw00}, with diameters of $\gapp75$ kpc \citep{gw00}, perhaps up to $\approx300$ kpc \citep{cwwa11}.

\subsubsection{Alternative gas distributions}

For completeness, we investigate several alternative density distributions for the gas. These are typically
profiles that arise from dynamical models, often applied to the dark matter content of a galaxy, although we
are interested in their effects when applied to the distribution of the CNM. The Jaffe profile \citep{jaf83} models the
distribution of light in a spherical galaxy as $n = n_0\,(r_{\rm s}/r)^2/4\,\pi\,(1 + r/r_{\rm s})^2$, where $r_{\rm s}$
is the radius which contains half the total emitted light. In this case Eq. \ref{eq1} becomes
\begin{eqnarray}
\int^{\infty}_{\nu_{_{\rm ion}}}\frac{L_{\nu}}{h\nu}\,d{\nu}  
=  4\pi\,\alpha_{\rm A}\,n_0^2\int^{r_{\rm ion}}_{0}\,\frac{r^2\,(r_{\rm s}/r)^4}{(4\,\pi)^2\,(1 + r/r_{\rm s})^4}\, dr 
\nonumber \\ = -\frac{4}{3}\pi\,\alpha_{\rm A}\,n_0^2\,r_{\rm s}^3\bigg[\frac{r_{\rm s}\,(3\,r_{\rm s}^3 + 22\,r_{\rm s}^2\,r+30\,r_{\rm s}\,r^2+12\,r^3)}{r\,(r_{\rm s}+r)^3}  \nonumber \\ \hspace{-6mm}- 12\ln(r_{\rm s}+r ) + 12\ln(r_{\rm s})\bigg]_0^{r_{\rm ion}}.
\label{eq_jaffe2} 
\end{eqnarray}
However, due to the ``cuspy'' nature of the distribution, 
solving this over these limits yields infinities and between any reasonable limits yields unreasonably large numbers, even when the 
approximation
$n = n_0\,(r_{\rm s}/r)^2/4\,\pi\,(1 + r/r_{\rm s})^2 \approx n_0\,(r_{\rm s}/r)^2/4\,\pi\,(r/r_{\rm s})^2 = (n_0/4\,\pi)(r_{\rm s}/r)^4$
is used.  

\begin{figure}
\centering 
\includegraphics[angle=270,scale=0.73]{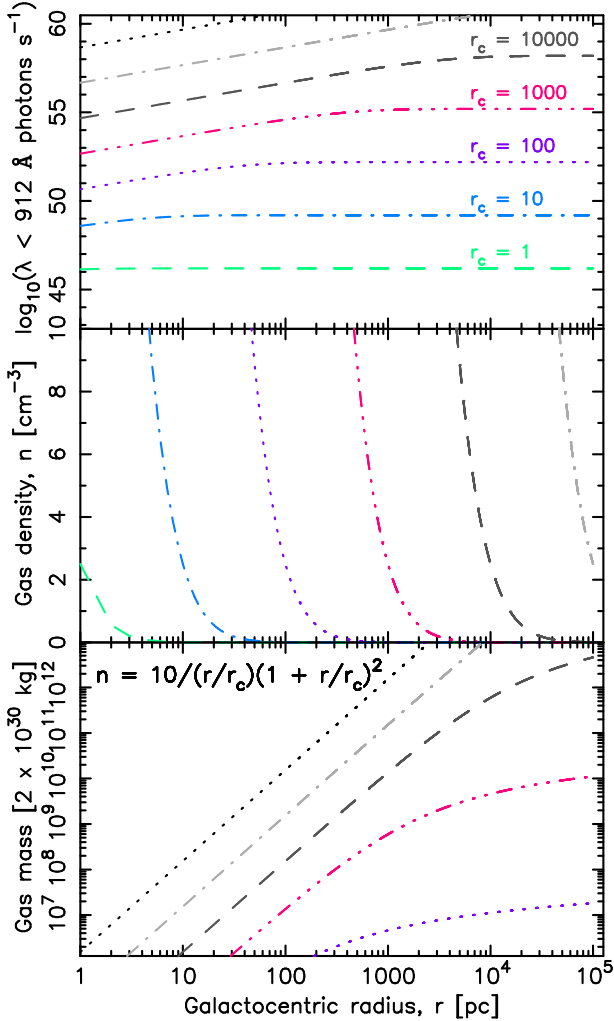}
\caption{As Fig. \ref{X-n} ($n_0 =10$ cm$^{-3}$ and $T =2000$ K) but for an NFW profile.}
\label{X-n-NFW}
\end{figure}

A similarly asymptotic density distribution is given by the Navarro--Frenk--White (NFW) profile \citep{nfw96}, which models 
the density as the distribution of dark matter in the halo, via $n = n_0\,(r_{\rm c}/r)/(1 + r/r_{\rm c})^2$, where $n_0 $ and $r_{\rm c}$ are the
core density and radius of the halo, respectively. Here the right hand side of Eq. \ref{eq1} becomes
\begin{eqnarray}
4\pi\,\alpha_{\rm A}\int^{r_{\rm ion}}_{0}\,n^2\,r^2 =  4\pi\,\alpha_{\rm A}\,n_0^2\int^{r_{\rm ion}}_{0}\,\frac{r^2\,(r_{\rm c}/r)^2}{(1 + r/r_{\rm c})^4}\, dr \nonumber \\= \frac{4}{3}\pi\,\alpha_{\rm A}\,n_0^2\,r_{\rm c}^3\left[1 - \frac{r_{\rm c}^3}{r_{\rm c}^3 + r_{\rm ion}}\right].
\label{eqnfw}
\end{eqnarray}
Unlike the Jaffe profile, the photoionsation equilibrium equation can be solved, again giving a ceiling luminosity, albeit less pronounced than 
for the exponential gas distribution (Fig. \ref{X-n-NFW}, top panel). The $L_{\rm UV} = 10^{23}$ \WpHz\ threshold of 
$2.9\times10^{56}$  $\lambda\leq 912$ \AA\  photons per second
gives $r_{\rm c}=2.6$ kpc, which is very close to the scale-length of the exponential disk and typical of that found for near-by galaxies \citep{dwb+08,odb+11}. 
For a spherical mass distribution the NFW profile gives
\begin{eqnarray}
M_{\rm gas} = 4\pi\,n_0\, \int_0^{r}\frac{r_{\rm c}/r}{(1 + r/r_{\rm c})^2}\,r^2\,dr 
\nonumber \\= 4\pi\,n_0\,r_{\rm c}^3\left[\frac{r_{\rm c}}{r_{\rm c} + r} -1 + \ln(r_{\rm c} + r) + \ln(r_{\rm c})\right],
\end{eqnarray}
from which $r_{\rm c}=2.6$ kpc gives a total gas mass of $M_{\rm gas} =7.1\times10^{11}$ \Mo, which, not surprisingly
given the distribution used, is close to the expected value for the dynamical mass.

\begin{figure}
\centering 
\includegraphics[angle=270,scale=0.73]{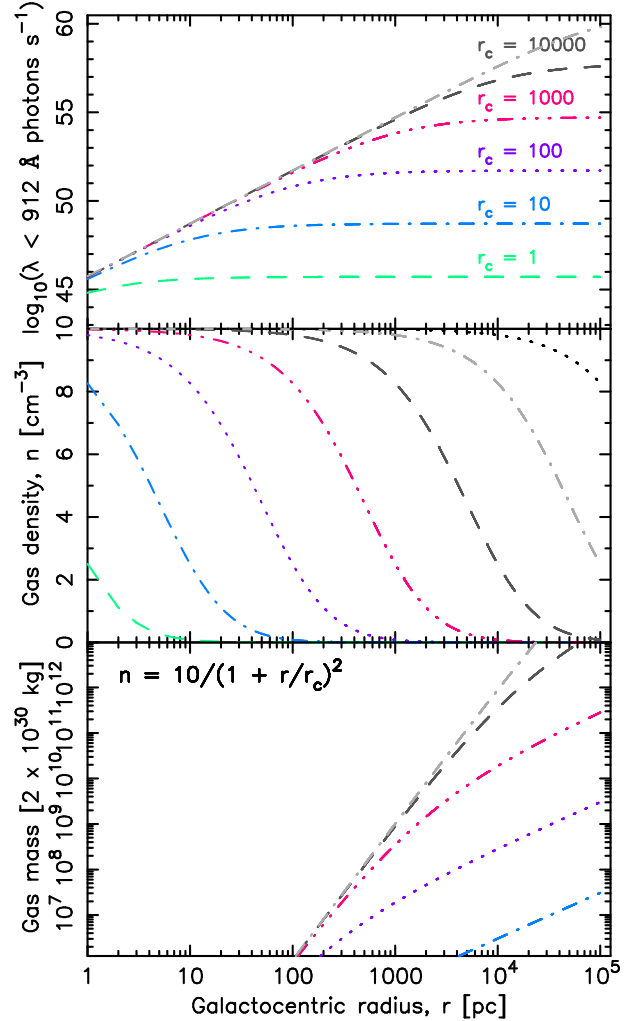}
\caption{As Fig. \ref{X-n-NFW} ($n_0 =10$ cm$^{-3}$ and $T =2000$ K) but for a halo profile with an isothermal sphere density distribution.}
\label{X-n-NFW-iso}
\end{figure}

The asymptotic density distribution of the NFW profile can be avoided by employing the halo density distribution of an isothermal sphere
\citep{bbs91}, i.e. $n = n_0/(1 + r/r_{\rm c})^2$, Fig. \ref{X-n-NFW-iso} (middle panel).
For this, the right hand side of Eq. \ref{eq1} gives
\begin{eqnarray}
4\pi\,\alpha_{\rm A}\int^{r_{\rm ion}}_{0}\,n^2\,r^2 =  4\pi\,\alpha_{\rm A}\,n_0^2\int^{r_{\rm ion}}_{0}\,\frac{r^2}{(1 + r/r_{\rm c})^4}\, dr \nonumber \\ = \frac{4}{3}\pi\,\alpha_{\rm A}\,n_0^2\,r_{\rm c}^3\left[1 - \frac{r_{\rm c}(r_{\rm c}^2 + 3\,r_{\rm c}\,r + 3\,r^2)}{(r_{\rm c} + r)^3}\right],
\label{eqiso}
\end{eqnarray}
which yields $r_{\rm c}\approx3.7$ kpc for $2.9\times10^{56}$  $\lambda\leq 912$ \AA\  photons per second. Note that 
the ceiling luminosities are more pronounced than for the NFW profile, due to
the steeper power law rise at $r < r_{\rm c}$, similar to that of the exponential gas distribution
(Fig. \ref{X-n-NFW-iso}, top panel).  For a spherical
mass distribution the profile gives
\begin{eqnarray}
M_{\rm gas} = 4\pi\,n_0\, \int_0^{r}\frac{r^2}{(1 + r/r_{\rm c})^2}\,dr 
\nonumber \\= 4\pi\,n_0\,r_{\rm c}^3\left[1 + \frac{r}{r_{\rm c}}  - \frac{r_{\rm c}}{r_{\rm c} + r} - 2\ln(r_{\rm c} + r) + 2\ln(r_{\rm c})\right], \nonumber \\
\end{eqnarray}
which again gives an physically unrealistic total gas mass, $M_{\rm gas} \approx9\times10^{16}$ \Mo.

Based on the derived gas masses, we therefore conclude that the exponential distribution is the most relevant
to the density profile of the CNM within the host galaxy, although the alternative models do reproduce the observed 
critical $\lambda\leq 912$ \AA\ luminosity for a similar scale-length (all of which, at $\approx3$ kpc, are close to that of the Milky Way). 
Although the exponential distribution may apply to our own galaxy \citep{kdkh07} and, through the assumption of the canonical values,
reproduces the observed critical luminosity and gas mass, currently we do not have the means to map the density profile of the CNM
at high redshift. Saying this, the important point is that all of these distributions introduces a critical luminosity
above which all of the gas is ionised: At lower radii,  when
the last term in Eqs. \ref{eq2}, \ref{eqnfw} and \ref{eqiso} are important, a power law is seen where $\log_{10} r_{\rm ion}\propto \log_{10}
\int({L_{\nu}}/{h\nu})\,d{\nu}$, which is the also the case for a constant gas density profile \citep{ost89}. However,
for a sufficiently high luminosity, this term is negligible and $r_{\rm ion} \rightarrow \infty$, where the decreasing 
UV flux with distance from the AGN maintains a sufficient number photons to fully ionise the thinning gas. 

\section{Discussion and Conclusions}

From the first high redshift survey of associated 21-cm absorption, 
\citet{cww+08} found an apparent lack of cool neutral gas within the hosts of radio galaxies and quasars. Upon an
analysis of the photometry of each source, a strong correlation between the
$\lambda = 1216$ \AA\ 
continuum luminosity of the AGN and the non-detection of 21-cm was found.  Specifically, that 21-cm
absorption has never been detected in a source in which the luminosity exceeds $L_{\rm 1216} \sim10^{23}$ \WpHz. Although
other factors may contribute to the raising of the spin temperature of the gas, the fact that 21-cm cannot be detected
above the ground state, in conjunction with the lack of detections above this critical luminosity (significant at
$4.76\sigma$), strongly suggests that excitation by $\lambda \leq1216$ \AA\ photons is the
dominant cause of the dearth of 21-cm in optically bright radio sources.

Here we demonstrate that this critical luminosity is also applicable to ionising ($\lambda \leq912$ \AA) photons, showing that
associated 21-cm is not detected for any source where $L_{912} \gapp10^{23}$ \WpHz\ or, more precisely, when there
are $\gapp2.9\times10^{56}$ ionising photons sec$^{-1}$. 
Applying this photoionsation rate, together with various gas density distribution models to the equation of photoionsation equilibrium,
from canonical values for the gas density
($n_0 = 10$ cm$^{-3}$) and the recombination rate coefficient ($\alpha_{\rm A}=1.27\times10^{-12}$ cm$^3$ sec$^{-1}$):
\begin{itemize}
\item We obtain the observed critical photon rate for a
  scale-length of $\approx3$ kpc for all of the tested profiles (exponential, NFW and isothermal sphere). This 
  scale-length is the same as that for the \HI\
  in the Milky Way, thus suggesting that the observed critical value is just sufficient to ionise all of the
  neutral gas within a large spiral galaxy.

\item This scale-length gives:
  \begin{itemize}
     \item For an exponential distribution within a disk,  a total gas mass of $M_{\rm gas} =7.5\times10^{9}$ \Mo, typical of that found from 21-cm emission studies 
       of low redshift galaxies.
       \item For the NFW and isothermal sphere distributions, a total gas mass which exceeds the total expected dynamical mass of the galaxy.
\end{itemize}
\end{itemize}
This leads us to conclude that the exponential profile is the more applicable to the distribution of the CNM (see \citealt{kk09}), although
all of the models give a critical UV luminosity. That is,  for a gas profile in which the density decreases with
distance from the nucleus, the Str\"{o}mgren sphere has an infinite radius for a finite luminosity. This suggests
that a balance is maintained between the decreasing number of photons and number of particles with increasing
distance from the ionising source.

For the sources under consideration here, the critical photon rate (where $L_{912} \sim L_{1216} \sim 10^{23}$ \WpHz)
is consistent with the dearth of 21-cm detections 
in all searched high redshift sources \citep{cww+08}. A ``proximity effect''   for highly ionised Lyman-\AL\ forest clouds 
has previously been noted \citep{wcs81,bdo88}, where in these intervening systems the high ionising flux from the QSO is believed to be
responsible for the decrease in the number density of the Lyman-\AL\ lines as the redshift of the absorbing galaxy
approaches that of the QSO ($z_{\rm abs}\rightarrow z_{\rm QSO}$).\footnote{\citet{be69} also show that both the 21-cm  and the Lyman-\AL\ flux can contribute to higher spin temperatures at absorber--quasar separations of less than a few tens of kpc.} 

However, until our high redshift survey of radio galaxies and quasars, no such effect was known for the 21-cm
transition.\footnote{From a study of absorber clustering around QSOs in the SDSS DR3, \citet{wkw+08} suggest that
  the QSO destroys the Mg{\sc \,ii} clouds out to beyond 800 kpc. Mg{\sc \,ii} has an ionisation potential of 15.04 eV,
close that of \HI\ (13.60 eV) and so a similar critical $\lambda \leq827$ \AA\ luminosity could perhaps account for this.}
Furthermore, the 21-cm effect is striking in that, rather than a gradual decrease in associated 21-cm absorption with
increasing ultra-violet luminosity, there is an abrupt cut-off in the 21-cm detection rate at a single critical
luminosity.  We show here that, for a typical spectral energy distribution, this is the luminosity required to completely ionise the large-scale
distribution of atomic gas in a large spiral galaxy.\footnote{Although not searched in 21-cm, we derive an ionising photon
  rate of $3.6\times10^{57}$ sec$^{-1}$ ($L_{\rm 912} = 2.1\times10^{24}$ \WpHz) for PKS\,0424--131. Since this is
  an order of magnitude higher than the critical rate, it could account for ``the mysterious absence of neutral hydrogen'', as traced by
  Lyman-\AL\ emission, close to this source \citep{fb04}. \citet{bwbd12} also suggest that the UV emission could be
  responsible, thus suppressing the star formation.}

There is the possibility that the non-detection of 21-cm absorption in the high-luminosity sample is simply due to orientation effects -- as mentioned previously
(Sect.\ref{intro}),  the majority (17 of 19) of the $L_{\rm UV} \gapp10^{23}$ \WpHz\ sources are type-1 AGN.
In the plane of the torus there may exist large columns of neutral gas, shielded from the radiation 
\citep{hp07}.  The orientation of these objects could explain their high UV luminosities, however there are also many
type-1 objects below $L_{\rm UV} \lapp10^{23}$ \WpHz, which exhibit a 50\% probability
of detecting 21-cm absorption. This suggests that the absorbing gas is located in the large-scale disk, which must be
randomly oriented with respect to the torus \citep{sksa97,nw99,cw10,le10}. 

Therefore, if the non-detection of absorption in
the $L_{\rm UV} \gapp10^{23}$ \WpHz\ sources is due to line-of-sight effects, these differ from the lower luminosity
AGN in that, unlike these, the large-scale absorbing disk is always aligned with the obscuring torus. 
It is possible that for these, the high UV luminosities are indeed the consequence of a direct view to the AGN
unobscured by the large-scale disk. This would also explain the lower luminosity type-1 objects in which 21-cm is
detected, in that the absorbing gas attenuates the radiation. 

Using the orientation as the sole explanation for the differences in absorption rates can not, however, explain why there
is a critical value in the UV luminosities rather than a gradual decrease in absorption rates as the luminosity increases. The 
model presented here naturally yields a critical luminosity, and we have shown that this value is sufficient to ionise all of the CNM in a large
spiral galaxy.

Our model is supported by the observation of greatly-decreased $250\mu$m emission from AGN above a critical luminosity \citep{psv+12}. 
The emission is not restricted to the line-of-sight to the nucleus, indicating that the entire galaxy is affected by the presence of 
a luminous AGN.


Finding redshifted 21-cm absorption is a major science goal of the Square Kilometre Array (SKA). 
The results found here suggest that the lack of cool neutral gas is not due to a sensitivity issue with current radio
telescopes, but a real effect caused by the presence of a quasar in a galaxy of gas.  That is, {\em the known high
  redshift radio galaxies and quasars 
are probably devoid 
of a large-scale distribution of neutral  gas.}\protect{\footnote{Simulations suggest that,
  while the \HI\ content in galaxies is similar to present day values, the molecular component (\MOLH) is larger at
  $z\gapp1$ \citep{or09a}. For gas densities typical of molecular clouds, $n_0 \sim1000$ cm$^{-3}$, the critical
  ionising photon flux of $2.9\times10^{56}$ sec$^{-1}$ gives a scale-length of $\approx100$ pc at 2000~K, or
  $\approx50$ pc at a more realistic 20 K. Conversely, luminosities of $L_{\rm 912}\approx2\times10^{27}$ \WpHz\
  ($3\times10^{60}$ photons sec$^{-1}$) and $4\times10^{28}$ \WpHz\ ($5\times10^{61}$ photons sec$^{-1}$) are required
  to fully ionise a CNM with $n_0 \sim1000$ cm$^{-3}$ with a scale-length of 3 kpc at 2000 and 20~K, respectively. If
  the deficit of neutral gas at high redshift was due to the gas being mostly molecular, there would be little reason
  for a lack of 21-cm to be correlated with the $\lambda = 912$ \AA\ luminosity (although a lack of \MOLH\ could be). In
  any case, the \HI\ deficit is observed at all redshifts, as well as there being no observational evidence for a high
  molecular gas content within the hosts of radio galaxies and quasars \citep{cwc+11}.}} 
This neutral gas provides the fuel for star formation, and our result suggests that the AGN could therefore
suppress star formation in the large-scale disk. Note also that, although the 250 $\mu$m fluxes are also subject to a 
sensitivity limit, the deficit of 250 $\mu$m above a
critical luminosity is further evidence of a suppression of star formation \citep{psv+12}. Here we show that this is not
a sensitivity issue, but that the neutral gas, and most likely, appreciable star formation activity is simply not present.

Therefore, even the SKA will be unlikely to find this cool gas in the objects currently known.
Where it will excel, however, is in blind surveys of radio sources from which the
visible light is too faint to be detected by optical instruments \citep{cww09}. 
Although unseen, these sources must exist in order to have had star formation within the
host galaxies of early AGN. 
As such, the traditional optical selection of targets must
be abandoned in order to find the missing star-forming material within high redshift radio sources.\\

We would like to thank John Webb, Bob Carswell, Elliot Koch, Julian Berengut, Catherine Greenhill and Nigel Badnell 
for their helpful input and advice.  
This research has made use of the NASA/IPAC Extragalactic Database
(NED) which is operated by the Jet Propulsion Laboratory, California
Institute of Technology, under contract with the National Aeronautics
and Space Administration. 
This research has
also made use of NASA's Astrophysics Data System Bibliographic
Service.
The Centre for All-sky Astrophysics is an Australian Research Council Centre of Excellence, funded by grant CE110001020.


\end{document}